# Identical Photons from Multiple Tin-Vacancy Centers in Diamond

*Yasuyuki Narita[1], Peng Wang[1], Kazuki Oba[1], Yoshiyuki Miyamoto[2], Takashi Taniguchi[3], Shinobu Onoda[4], Mutsuko Hatano[1], Takayuki Iwasaki[1,*]*

[1]Department of Electrical and Electronic Engineering, School of Engineering, Tokyo Institute of Technology, Meguro, Tokyo 152-8552, Japan

[2]Research Center for Computational Design of Advanced Functional Materials, National Institute of Advanced Industrial Science and Technology, Tsukuba, Ibaraki 305-8568, Japan

[3]International Center for Materials Nanoarchitectonics, National Institute for Materials Science, Tsukuba, Ibaraki 305-0044, Japan

[4]Takasaki Advanced Radiation Research Institute, National Institutes for Quantum Science and Technology, 1233 Watanuki, Takasaki, Gunma 370-1292, Japan

email: iwasaki.t.aj@m.titech.ac.jp

**Abstract**

We report the narrow inhomogeneous distribution of the zero-phonon line from tin-vacancy (SnV) centers in diamond and the overlap of spectra from multiple separated SnV centers. Photoluminescence excitation spectroscopy measurements at a cryogenic temperature showed that SnV centers exhibit stable fluorescence and linewidths close to the Fourier transform-limited linewidth. The inhomogeneous distribution was as low as ~4 GHz, which enabled the observation of Sn isotope-dependent resonant frequencies. Owing to the narrow inhomogeneous distribution, we observed multiple separated SnV centers showing identical photons with almost the same wavelength and linewidth. Identical SnV centers were also observed even in different diamond samples, confirming the reliable fabrication of the high-quality SnV centers.

Quantum networks require entanglement generation between distant quantum network nodes, including quantum repeaters.[1] Solid-state quantum emitters are a promising platform for quantum light-matter interfaces to construct quantum nodes.[2,3] The generation of quantum entanglement among distant nodes has been demonstrated using nitrogen-vacancy (NV) centers in diamond,[4,5] while the NV center has drawbacks in terms of optical properties: a low fraction of the zero-phonon line (ZPL) against the total emission and a fluorescence line that is subjected to external electric noise. Group-IV color centers in diamond, silicon-vacancy (SiV),[6–8] germanium-vacancy (GeV),[9,10] tin-vacancy (SnV),[11,12] and lead-vacancy (PbV)[13–15] centers, with inversion symmetry, can solve these problems.[16–19] In particular, SnV and PbV centers are interesting systems from the viewpoint of the spin property. They are expected to exhibit spin coherence times of milliseconds at Kelvin temperatures owing to the large spin-orbit interaction in the ground state. This is in contrast to the SiV center, which requires a temperature < 1 K in a dilution refrigerator to achieve a long spin coherence time.[20] A spin coherence time of 0.3 ms has been reported for a SnV center with all optical operations.[21] Observations of Fourier transform-limited linewidth (FTL),[22,23] incorporation in nanocavities,[24–26] and two-photon interference using one SnV center in a nanostructured diamond[27] have also been demonstrated, making the SnV center promising as a light-matter interface. However, the generation of photons with identical linewidths and fluorescence wavelengths from multiple separated SnV centers has yet to be reported, which would play a key role in generating remote entanglement[28] based on the two-photon interference[29,30] between quantum network nodes. Control of the linewidth and wavelength is challenging in solid-state materials, especially for a SnV center composed of heavy atoms, because its incorporation into diamond causes more defects and higher strain around the emitters. In this study, we observed identical photons from multiple SnV centers formed by ion implantation and subsequent high-temperature annealing at 2100 °C under a high pressure of 7.7 GPa. The SnV centers were found to be stable over time and showed narrow linewidths close to the FTL. Furthermore, statistical measurements revealed a narrow inhomogeneous distribution, in which we found Sn-isotope-dependent fluorescence energy. Finally, the narrow inhomogeneous distribution enabled us to find out multiple SnV centers emitting identical photons.

**Results and Discussion**

We fabricated SnV centers in (001) diamond substrates by high-energy ion implantation at an acceleration energy of 18 MeV and subsequent HPHT annealing. Two samples, hereinafter called Sample 1 and Sample 2, were measured. Unless otherwise noted, the experimental results are obtained using Sample 1. A SnV center takes a split-vacancy configuration with an interstitial Sn atom between two carbon vacancies (Fig. 1a). This structure corresponds to the $D_{3d}$ symmetry, in which no permanent electric dipole is present. A negatively charged SnV center forms the S=1/2 system. The energy levels in the ground and excited states are split due to the spin-orbit interaction, as shown in Fig. 1b. Thus, four optical transitions (A-D) occur upon non-resonant excitation. Figure 1c shows a PL spectrum from ensemble SnV centers at ~6 K. The two sharp lines are the C- and D-peaks emitted from the lower level in the excited state to the two ground states. The energy difference corresponding to the ground state splitting is 821 GHz, in agreement with a previous report.[11] The A- and B-peaks are not visible because the thermal energy does not pump electrons to the upper level in the excited state at this temperature. It is worth noting that only one peak can be observed for each of the C- and D-peaks, even for the measurement of the ensemble

SnV centers, indicating that the defects created during ion implantation are efficiently recovered by the HPHT annealing.

Figure 1d shows a confocal plane scan mapping obtained using resonant laser excitation. The phonon sideband (PSB) was detected during the scan. We see multiple isolated bright spots that are expected to have very close absorption wavelengths. The depth scan in Fig. 1e revealed that the bright spots are located at approximately 1.2 μm in depth from the surface, indicated as a dashed line. The correction considering the mismatched refractive index between vacuum and diamond[31] estimates an actual depth of ~3.2 μm (Supporting Information). This agrees with the projected depth obtained by the SRIM calculation of the implanted Sn ions.

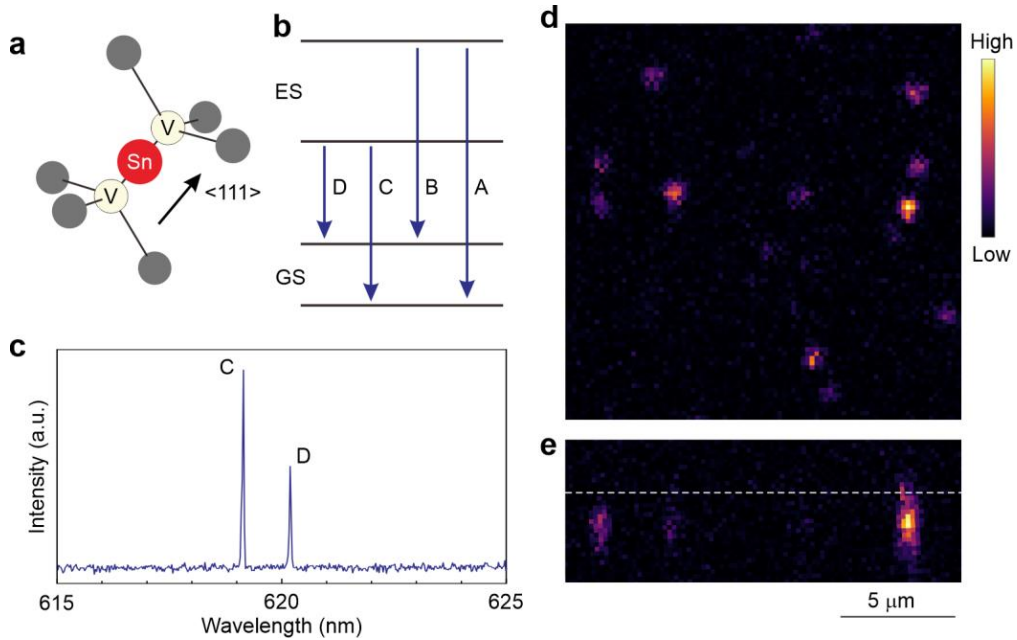

**Figure 1.** Structure and optical properties of the SnV center. (a) Atomic structure. The main axis is along the <111> direction. (b) Energy level with a split ground state (GS) and excited state (ES). (c) PL spectrum at ~6 K. Only two lines, called C- and D-peaks, are observed due to the low

electron occupation in the upper level of the excited state at low temperature. (d) Plane and (e) depth confocal scan mappings obtained using a resonant laser on the C-peak. The dashed line in Panel (e) indicates the diamond surface, estimated by laser reflection.

The stability of the resonant peaks over time was investigated by repeating the scanning of the laser wavelength around the C-peak at one laser spot. Figure 2a shows consecutively measured photoluminescence excitation (PLE) spectra for a duration of approximately 12 min. Here, we used a low laser power of 0.8 nW to avoid termination of the fluorescence due to charge conversion.[22] At this laser power, fluorescence termination was not observed even without irradiation from another non-resonant laser, which was utilized for charge stability.[32] At the measured spot, three bright lines are clearly observed with a frequency separation of ~400 MHz. These lines correspond to different emitters with energy levels slightly modified by the lattice strain. The linewidths of the three peaks are 32, 33, and 32 MHz (from low to high detuning). The excited lifetime of the SnV center has been reported to be 4–8 ns,[11,22,23,27,33] corresponding to an FTL of 20–40 MHz. Therefore, the observed linewidths in this study are comparable to the reported FTL. Importantly, the peak positions do not show notable fluctuation, evidencing the very high stability of the SnV emission, as expected for the $D_{3d}$ symmetry with a vanishing permanent electric dipole. The stability of the resonant peak was also examined at another position (Fig. 2b). Again, we see three sharp lines with narrow linewidths: 33 and 32 MHz for the peaks around zero detuning. The two lines even overlap at one confocal spot measurement. This suggests that the inhomogeneous distribution is strongly suppressed in the sample. Note that the gradual increase in the peak intensity of the right peak over time is caused by the drift of the measurement position.

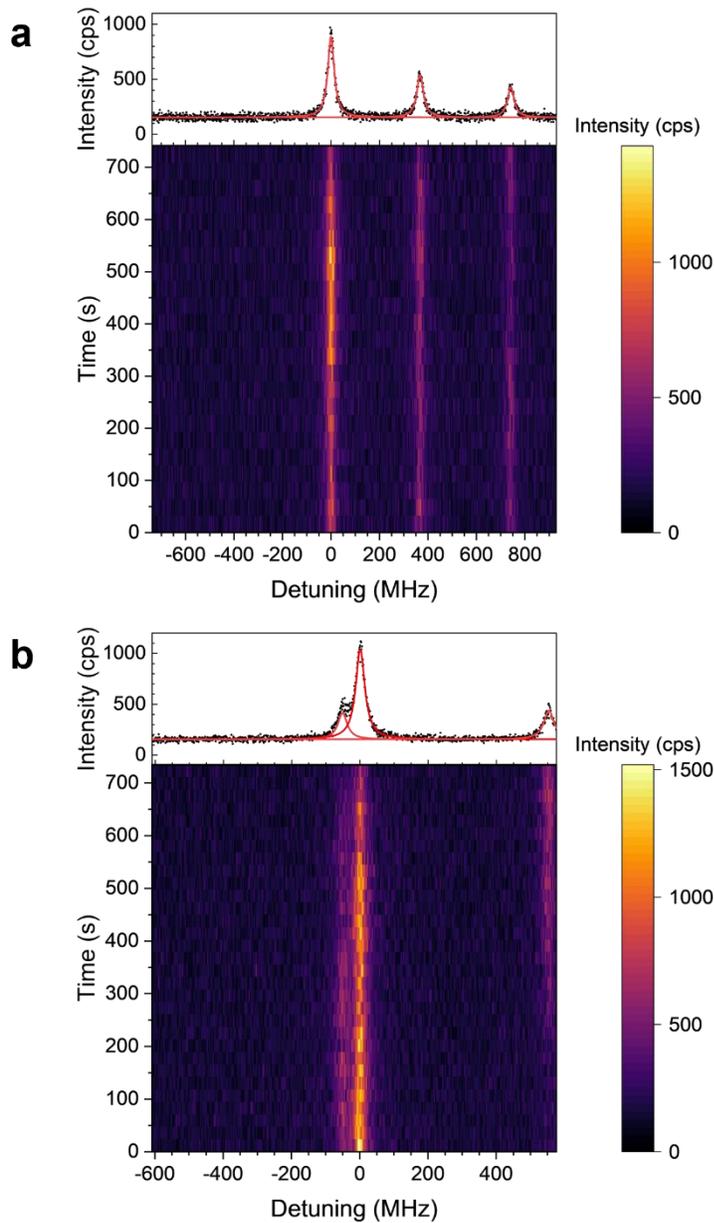

**Figure 2.** Stability of the SnV emission. The measurements were performed at two different laser spots. The bottom panels show the stability of the resonant peaks over time. The data in (a) and (b) include 23 and 31 scans, respectively. The PLE spectra in the top panels are the averaged curves of the bottom panels. The red curves are fitted.

The inhomogeneous distribution of the formed SnV centers was investigated by performing wide-range PLE scans (47 GHz) at two areas in the sample, denoted as areas 1 and 2, for a total of 43 spots. Each area was approximately $20 \times 20$ μm$^2$ near the center of the sample. Figure 4a shows the histogram of the C-transition of 160 SnV centers. We clearly see three regions in the histogram, hereinafter called P1, P2, and P3. A large portion of SnV centers exhibit energy in the P1 region. The three regions are thought to result from the energy shift due to the Sn isotopes. As shown for SiV and GeV centers,[10,34–36] the isotope of the group-IV element causes a shift in the ZPL according to the mass difference. For the preparation of the sample evaluated here, the ion implantation conditions were set to predominantly select $^{120}$Sn ions rather than using the natural abundance. Thus, the P1 regime has the highest count because of the fluorescence from the $^{120}$SnV centers. However, the other isotopes could be contaminated during ion implantation due to the imperfect isolation of the target ion resulting from the heaviness of the Sn ions.

Assuming that the local vibration modes of the Sn atom determine the ZPL energy, the energy shift between $^n$SnV and $^{n^*}$SnV centers with different masses $m_n$ and $m_{n^*}$ ($m_n < m_{n^*}$) is expressed by the following equation:[34,37]

$$\Delta E_{n,n^*} = E_n - E_{n^*} = \frac{\hbar}{2}\left(\sum \Omega^e_{n_{S_n}} - \sum \Omega^g_{n_{S_n}}\right) \times \left(1 - \sqrt{\frac{m_n}{m_{n^*}}}\right) \quad (1)$$

where $E_n$ and $E_{n^*}$ are the ZPL energies of the $^n$SnV and $^{n^*}$SnV centers, respectively. $\Omega^e_{n_{S_n}}$ and $\Omega^g_{n_{S_n}}$ are the vibration frequencies of $^n$SnV in the excited and ground states, respectively. The phonon frequencies include the Sn vibrations in both the axial (A$_{2u}$) and transverse (E$_u$) directions.[37] It has been reported that lighter Si and Ge isotopes lead to the shift of ZPL to higher energies.[34,35] As demonstrated theoretically, this is because the electron is optically excited from a less localized state to a localized state, leading to an increase in the vibration frequency in the excited state.[37,38] This makes the difference in the vibration frequencies in Equation (1) positive, resulting in $\Delta E_{n,n^*} > 0$. Our calculations suggest a higher vibration frequency in the excited state than in the ground state for the SnV center as well (Supporting Information). Thus, $\Delta E_{n,n^*}$ takes a positive value, and accordingly, the SnV center composed of the lighter Sn isotope shows a higher energy. This means that P2 and P3 with higher energies are thought to correspond to lighter isotopes, i.e., $^{119}$SnV and $^{118}$SnV centers, respectively. From Equation (1), $\Delta E_{118,119}$ and $\Delta E_{118,120}$ are expressed as

$$\Delta E_{118,119} = \frac{\hbar}{2}\left(\sum \Omega^e_{118_{S_n}} - \sum \Omega^g_{118_{S_n}}\right) \times \left(1 - \sqrt{\frac{m_{118}}{m_{119}}}\right) \quad (2)$$

$$\Delta E_{118,120} = \frac{\hbar}{2}\left(\sum \Omega^e_{118_{S_n}} - \sum \Omega^g_{118_{S_n}}\right) \times \left(1 - \sqrt{\frac{m_{118}}{m_{120}}}\right) \quad (3)$$

Accordingly, the ratio of the ZPL shift becomes

$$\frac{\Delta E_{118,119}}{\Delta E_{118,120}} = \frac{1 - \sqrt{\frac{m_{118}}{m_{119}}}}{1 - \sqrt{\frac{m_{118}}{m_{120}}}} = 0.5 \quad (4)$$

From the Gaussian fitting of the histogram in Fig. 3a, $\Delta E_{118,119}$ and $\Delta E_{118,120}$ are estimated to be 10.9 GHz and 17.9 GHz, respectively. These values lead to a ratio of the ZPL shift ($\Delta E_{118,119}/\Delta E_{118,120}$) of ~0.6, which roughly agrees with the theoretical ratio. The natural abundance of the Sn isotopes close to $^{120}$Sn is $^{117}$Sn : $^{118}$Sn : $^{119}$Sn : $^{120}$Sn : $^{122}$Sn = 7.7% : 24.2% :

8.6% : 32.6% : 4.6%.[39] While $^{120}$Sn ions were dominantly implanted in this study, $^{119}$Sn has a high possibility of contaminating the sample, leading to the P2 region in Fig. 3a. Although $^{118}$Sn is two masses different from $^{120}$Sn, its high natural abundance causes the P3 region. The linewidth of P3 is clearly broader than those of the other two regions, which possibly originates from residual strain in the diamond lattice. This might lead to the discrepancy of the ZPL shift ($\Delta E_{118,119}/\Delta E_{118,120}$) from the theory. Note that we observed a few peaks in the higher positive detuning outside of the region in Fig. 3a during 10 scans at different spots, which may also appear due to strain. It is worth stressing that the isotope-distinguished observation of $^{119}$SnV is important for utilizing a long-lived nuclear spin for quantum memory.[40–42]

Here, we discuss the isotopic energy shift depending on the group-IV element. Experimentally, the energy shifts against the unit isotopic mass change are 87, 15, and 7 GHz for SiV,[34] GeV,[35] and SnV centers (between $^{119}$SnV and $^{120}$SnV in this study), respectively. The heavier group-IV emitter shows a smaller isotopic shift. The calculation of the vibration frequencies of the A$_{2u}$ and E$_u$ modes of the group-IV impurity gives rise to the theoretical energy shift using Equation (1).[37] The value of the SiV center estimated by the first-principles calculation[37] is plotted in Fig. 3b. In this study, we calculated the vibration frequencies of the SiV and SnV centers by first-principles calculations, showing the decrease in the energy shift as a function of the mass of the group-IV elements (Fig. 3b). Another calculation[35] of the SiV and GeV centers shows a similar decreasing tendency. With the relation of $\Omega_{n_{S_n}} = \sqrt{k/m_n}$, where $k$ is the force constant, the energy shift against the unit isotopic mass change $\Delta E_{n,n+1}$ becomes proportional to $1/\sqrt{m_n} - 1/\sqrt{m_{n+1}}$ according to Equation (1). This relationship is shown as the dashed line in Fig. 3b, the slope of which agrees with the experiment and calculations. Thus, the decreasing tendency would dominantly originate from the group-IV impurity mass with modification of the force constant.

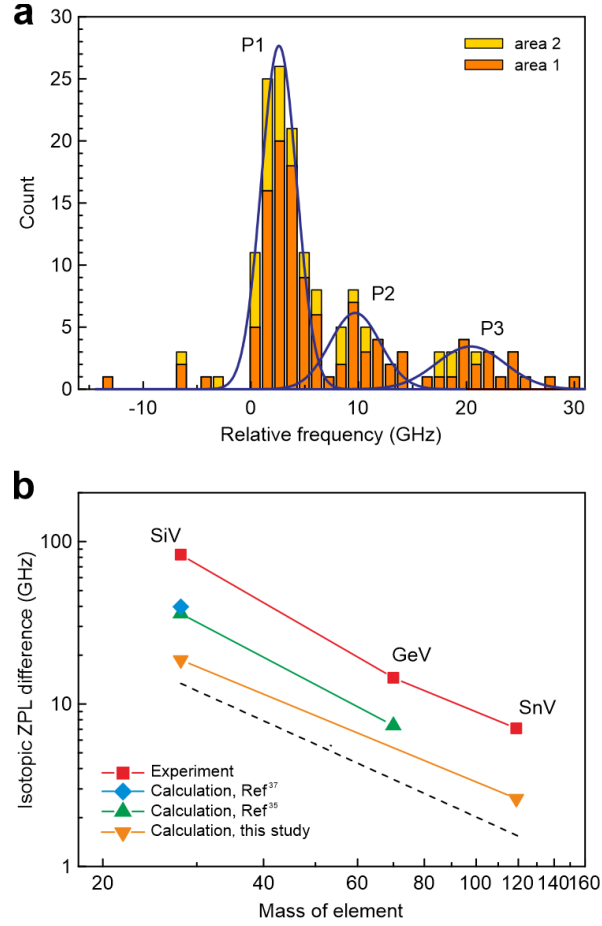

**Fig. 3.** Inhomogeneous distribution. (a) Histogram of the PLE resonant peaks of 160 SnV centers. The measurements were performed in two areas of Sample 1, and the data are stacked with the different colors. The frequency is given as the shift from 484.130 THz. The blue lines are the fitting curves for the three regions. The three peaks of P1, P2, and P3 are thought to correspond to $^{120}$SnV, $^{119}$SnV, and $^{118}$SnV, respectively. (b) Isotopic difference of the ZPL energy depending on the mass of the group-IV element. The experimental values of the SiV and GeV centers are obtained from previous reports.[34,35] The dashed line denotes the relation $\Delta E_{n,n+1} \propto 1/\sqrt{m_n} - 1/\sqrt{m_{n+1}}$.

The highest count group (P1 region) shows a narrow inhomogeneous distribution with an FWHM of 3.9 GHz, as shown in Fig. 3a. In the P1 energy range for area 1, the probability that two random emitters have fluorescence energies within 1 FTL (assumed to be 30 MHz) is estimated to be 1.1%. This is comparable to that of $V_{Si}$ in SiC.[43] We observed similar narrow inhomogeneous distributions at different areas in the sample. Furthermore, the reproducibility of the sample fabrication was also confirmed by observing the narrow inhomogeneous distribution of another sample, Sample 2, as shown in Fig. S3 in Supporting Information.

Figure 4a shows multiple SnV centers with a mutual overlap in the PLE spectra, which were found by performing resonant confocal scan.[7] In particular, the inner two emitters, ~10 μm apart, show identical photons with almost the same linewidths of 35 MHz (green curve) and 38 MHz (blue curve) and a very small energy difference of ~4 MHz, corresponding to 1/8 of FTL. Identical photons at different resonant energies are also shown in Fig. S2 in Supporting Information. We investigated Sample 2 and found a SnV center showing a resonant energy similar to those in Fig. 4a. Figure 4b depicts the spectral overlap of the emitters in Samples 1 and 2. This result further confirms the reliable fabrication of high-quality SnV quantum emitters. The generation of identical photons from the different samples will lead to the observation of two-photon interference between distant SnV centers.

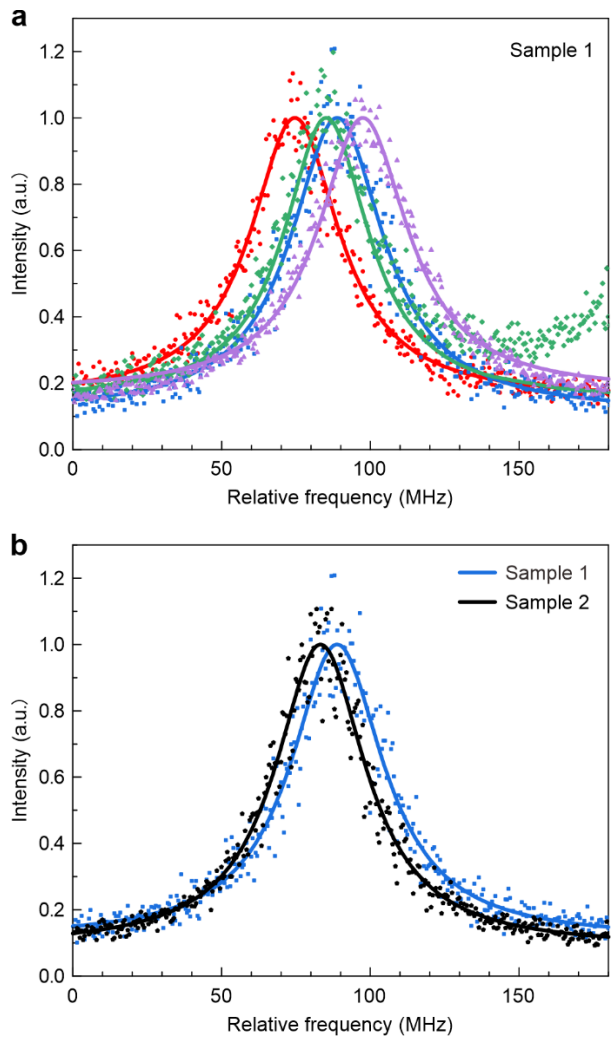

**Fig. 4.** Observation of identical photons from multiple SnV centers. (a) PLE spectra from four SnV emitters in Sample 1. (b) PLE spectra from two SnV emitters in Sample 1 and Sample 2. The blue spectrum in Sample 1 is the same as that in Panel (a). The experimental data are averaged for 3-6 laser scans. The solid lines are fitted. The data and fitting are normalized with the maximum of the fitting for each curve. The frequency is given relative to 484.1319 THz for both panels.

Among the group-IV color centers in diamond, several studies have demonstrated statistically narrow inhomogeneous distributions of SiV centers. Ion implantation with subsequent annealing has shown inhomogeneous distributions of ~15 GHz.[44,45] For SiV centers in diamond films grown by microwave plasma chemical vapor deposition (CVD),[7,34,46] narrower distributions were observed, with an FWHM of 0.7-8 GHz. SiV centers in nanodiamonds grown under HPHT also showed a narrow distribution of 6.8 GHz.[47] The CVD and HPHT formations lead to the spectral overlap of multiple SiV centers. This indicates that the direct doping of impurities during diamond growth can reduce strain and defects around the color centers. On the other hand, here, we successfully obtained a narrow inhomogeneous distribution as low as 3.9 GHz, enabling us to observe identical photons from multiple SnV centers formed by ion implantation and subsequent HPHT annealing, even for doping of heavy Sn impurities. It is worth noting that although we do not control the polarization of the fluorescence from the SnV centers, it can be adjusted by using polarization optics[7,27,48–50] for two-photon interference.

There are two key points in achieving a narrow inhomogeneous distribution of the SnV centers based on ion implantation. One is the high-temperature annealing at 2100 °C. Although the heaviness of the Sn ions causes a number of defects in the diamond lattice during ion implantation, the high-temperature treatment efficiently addresses the lattice damage. Consequently, the strain around the emitters is largely suppressed. Another point is ion implantation at a high energy. The SnV centers are formed at a depth of ~3 µm from the surface. This suppresses the effect of strain and charged defects at the surface, which potentially shifts the energy level of the emitters. The effect of the surface defects on the inhomogeneous distribution will be investigated in the future, e.g. by incorporating into nanostructures.[24–26,51,52]

The inhomogeneous distribution of 3.9 GHz obtained in this study is the narrowest among the SnV centers. Although it is much broader than 1 FTL, the deterministic overlap of multiple SnV centers can be achieved with the combination of spectral tuning by strain engineering[53–56] or Stark shift.[57,58] Strain engineering demonstrates a wide tuning range of over 100 GHz,[54] although it requires sophisticated nanofabrication techniques. As another method, voltage application to an emitter alone can control the center frequency by the Stark effect. Its tuning range has been thus far limited to ~2 GHz without significant line broadening,[57,58] but this will enable us to overlap approximately half of our SnV centers in the P1 region (Sample 1, area 1) with a narrow inhomogeneous distribution.

**Conclusion**

We fabricated SnV centers with a narrow inhomogeneous distribution of fluorescence energy. The SnV centers exhibited time-stable resonant energy over 10 min and linewidths close to the FTL. In the histogram of the resonant energy, three regimes were clearly observed, corresponding to the Sn isotopes. By performing computational calculations, we revealed that a lighter Sn isotope leads to a shift to a higher resonant energy. In the narrowest inhomogeneous distribution region, with an FWHM of 3.9 GHz, we observed multiple SnV centers emitting identical photons with overlapping PLE spectra. The formation of high-quality SnV centers directly leads to the observation of Hong-Ou-Mandel interference between distant emitters and promises further progress of SnV centers in diamond to establish a quantum-light matter interface.

**Methods**

**Sample fabrication.** The SnV centers were formed in IIa-type (001) single-crystal diamond substrates (Element Six) by ion implantation and subsequent high-temperature annealing. Ion implantation was performed at an acceleration energy of 18 MeV provided from a 3 MV Tandem accelerator. The mass separator was used to mainly extract $^{120}$Sn$^{5+}$ ions, and the extracted ion beam passing through the slits after the mass spectrometry magnet was implanted into the samples with a designed fluence of $5 \times 10^8$ cm$^{-2}$, leading to a projected depth of approximately 3 μm by SRIM.[59] Then, we conducted the high-temperature treatment for the formation of SnV centers[11] at 2100 °C under 7.7 GPa for 20 min using a belt-type high-pressure apparatus.[60] Although surface etching and/or diamond deposition frequently occur during the high-pressure and high-temperature (HPHT) process,[15,22] high-energy ion implantation suppresses these effects on the optical properties of the SnV centers. We prepared two samples, Sample 1 and Sample 2, using the same procedures described above, but the surface of Sample 2 was polished after the HPHT annealing (see Supporting Information).

**Optical measurements.** The optical properties were investigated by using a home-built confocal fluorescence microscope equipped with a cryostat (Montana Instruments). All measurements were conducted at a low temperature of ~6 K. A 515 nm green laser (Cobolt Fandango) was used for non-resonant excitation measurements. The laser was focused onto the sample through a 50 × objective lens with a numerical aperture (NA) of 0.95. The fluorescence from the SnV centers was directed to avalanche photodiodes or a spectrometer.

PLE measurements by resonant excitation were performed by using a dye tunable laser (Sirah Lasertechnik Matisse 2 DS) at laser powers of 0.8–1 nW. A wavemeter (Highfinesse WS8-30) with a resolution of 1 MHz was used to monitor the laser wavelength. The PSB of the SnV centers was detected during resonant excitation. Confocal fluorescence mapping and PLE spectra were recorded with the Qudi program.[61]

**Computational calculation.** The frequencies of the Sn atom with $A_{2u}$ ($E_u$) modes under the ground and excited states were computed from the force field on it by replacement in the <111> (<1$\bar{1}$0>) crystallographic direction of cubic diamond with an amount of 0.05 Bohr. To calculate the force field, density functional theory (DFT) calculations were performed by using PBE functionals,[62] norm-conserving pseudopotentials,[63] and conjugate-gradient plane wave schemes as described in a previous report.[64] The details of the computational conditions are explained in Supporting Information. We tested the current computational scheme to obtain vibrational energies for Si atoms in the SiV center and obtained lower values than those reported in previous work,[37] which may be due to the choice of the current functional of the DFT as well as the pseudopotentials being different from the reference values.[37]

**Notes**

The authors declare no competing financial interests.

ACKNOWLEDGMENTS

This work is supported by JSPS KAKENHI Grant Numbers JP22H04962 and JP22H00210, the Toray Science Foundation, the MEXT Quantum Leap Flagship Program (MEXT Q-LEAP) Grant Number JPMXS0118067395, and JST Moonshot R&D Grant Number JPMJMS2062.

REFERENCES

(1)  Kimble, H. J. The Quantum Internet. *Nature* **2008**, *453*, 1023–1030.

(2)  Atatüre, M.; Englund, D.; Vamivakas, N.; Lee, S. Y.; Wrachtrup, J. Material Platforms for Spin-Based Photonic Quantum Technologies. *Nat. Rev. Mater.* **2018**, *3*, 38–51.

(3)  Awschalom, D. D.; Hanson, R.; Wrachtrup, J.; Zhou, B. B. Quantum Technologies with Optically Interfaced Solid-State Spins. *Nat. Photonics* **2018**, *12*, 516–527.

(4)  Hensen, B.; Bernien, H.; Dreaú, A. E.; Reiserer, A.; Kalb, N.; Blok, M. S.; Ruitenberg, J.; Vermeulen, R. F. L.; Schouten, R. N.; Abellán, C.; Amaya, W.; Pruneri, V.; Mitchell, M. W.; Markham, M.; Twitchen, D. J.; Elkouss, D.; Wehner, S.; Taminiau, T. H.; Hanson, R. Loophole-Free Bell Inequality Violation Using Electron Spins Separated by 1.3 Kilometres. *Nature* **2015**, *526*, 682–686.

(5)  Pompili, M.; Hermans, S. L. N.; Baier, S.; Beukers, H. K. C.; Humphreys, P. C.; Schouten, R. N.; Vermeulen, R. F. L.; Tiggelman, M. J.; dos Santos Martins, L.; Dirkse, B.; Wehner, S.; Hanson, R. Realization of a Multinode Quantum Network of Remote Solid-State Qubits. *Science* **2021**, *372*, 259–264.

(6)  Clark, C. D.; Thomson, I. J.; Kanda, H.; Kiaawi, I.; Sittas, G.; Thomson, J. J. Silicon Defects in Diamond. *Phys. Rev. B* **1995**, *51*, 16681–16688.

(7)  Sipahigil, A.; Jahnke, K. D.; Rogers, L. J.; Teraji, T.; Isoya, J.; Zibrov, A. S.; Jelezko, F.; Lukin, M. D. Indistinguishable Photons from Separated Silicon-Vacancy Centers in Diamond. *Phys. Rev. Lett.* **2014**, *113*, 113602.

(8)  Hepp, C.; Müller, T.; Waselowski, V.; Becker, J. N.; Pingault, B.; Sternschulte, H.; Steinmüller-Nethl, D.; Gali, A.; Maze, J. R.; Atatüre, M.; Becher, C. Electronic Structure of the Silicon Vacancy Color Center in Diamond. *Phys. Rev. Lett.* **2014**, *112*, 036405.

(9)  Iwasaki, T.; Ishibashi, F.; Miyamoto, Y.; Doi, Y.; Kobayashi, S.; Miyazaki, T.; Tahara, K.; Jahnke, K. D.; Rogers, L. J.; Naydenov, B.; Jelezko, F.; Yamasaki, S.; Nagamachi, S.;

Inubushi, T.; Mizuochi, N.; Hatano, M. Germanium-Vacancy Single Color Centers in Diamond. *Sci. Rep.* **2015**, *5*, 12882.

(10) Palyanov, Y. N.; Kupriyanov, I. N.; Borzdov, Y. M.; Surovtsev, N. V. Germanium: A New Catalyst for Diamond Synthesis and a New Optically Active Impurity in Diamond. *Sci. Rep.* **2015**, *5*, 14789.

(11) Iwasaki, T.; Miyamoto, Y.; Taniguchi, T.; Siyushev, P.; Metsch, M. H.; Jelezko, F.; Hatano, M. Tin-Vacancy Quantum Emitters in Diamond. *Phys. Rev. Lett.* **2017**, *119*, 253601.

(12) Tchernij, S. D.; Herzig, T.; Forneris, J.; Küpper, J.; Pezzagna, S.; Traina, P.; Moreva, E.; Degiovanni, I. P.; Brida, G.; Skukan, N.; Genovese, M.; Jakšić, M.; Meijer, J.; Olivero, P. Single-Photon-Emitting Optical Centers in Diamond Fabricated upon Sn Implantation. *ACS Photonics* **2017**, *4*, 2580–2586.

(13) Ditalia Tchernij, S.; Lühmann, T.; Herzig, T.; Küpper, J.; Damin, A.; Santonocito, S.; Signorile, M.; Traina, P.; Moreva, E.; Celegato, F.; Pezzagna, S.; Degiovanni, I. P.; Olivero, P.; Jakšić, M.; Meijer, J.; Genovese, P. M.; Forneris, J. Single-Photon Emitters in Lead-Implanted Single-Crystal Diamond. *ACS Photonics* **2018**, *5*, 4864–4871.

(14) Trusheim, M. E.; Wan, N. H.; Chen, K. C.; Ciccarino, C. J.; Flick, J.; Sundararaman, R.; Malladi, G.; Bersin, E.; Walsh, M.; Lienhard, B.; Bakhru, H.; Narang, P.; Englund, D. Lead-Related Quantum Emitters in Diamond. *Phys. Rev. B* **2019**, *99*, 075430.

(15) Wang, P.; Taniguchi, T.; Miyamoto, Y.; Hatano, M.; Iwasaki, T. Low-Temperature Spectroscopic Investigation of Lead-Vacancy Centers in Diamond Fabricated by High-Pressure and High-Temperature Treatment. *ACS Photonics* **2021**, *8*, 2947–2954.

(16) Iwasaki, T. Color Centers Based on Heavy Group-IV Elements. *Semicond. Semimetals* **2020**, *103*, 237–256.

(17) Ruf, M.; Wan, N. H.; Choi, H.; Englund, D.; Hanson, R. Quantum Networks Based on Color Centers in Diamond. *J. Appl. Phys.* **2021**, *130*, 070901.

(18) Bradac, C.; Gao, W.; Forneris, J.; Trusheim, M. E.; Aharonovich, I. Quantum Nanophotonics with Group IV Defects in Diamond. *Nat. Commun.* **2019**, *10*, 5625.

(19) Chen, D.; Zheludev, N.; Gao, W. bo. Building Blocks for Quantum Network Based on Group-IV Split-Vacancy Centers in Diamond. *Adv. Quantum Technol.* **2020**, *3*, 1900069.

(20) Sukachev, D. D.; Sipahigil, A.; Nguyen, C. T.; Bhaskar, M. K.; Evans, R. E.; Jelezko, F.; Lukin, M. D. Silicon-Vacancy Spin Qubit in Diamond: A Quantum Memory Exceeding 10 ms with Single-Shot State Readout. *Phys. Rev. Lett.* **2017**, *119*, 223602.

(21) Debroux, R.; Michaels, C. P.; Purser, C. M.; Wan, N.; Trusheim, M. E.; Arjona Martínez, J.; Parker, R. A.; Stramma, A. M.; Chen, K. C.; De Santis, L.; Alexeev, E. M.; Ferrari, A. C.; Englund, D.; Gangloff, D. A.; Atatüre, M. Quantum Control of the Tin-Vacancy Spin Qubit in Diamond. *Phys. Rev. X* **2021**, *11*, 041041.

(22) Görlitz, J.; Herrmann, D.; Thiering, G.; Fuchs, P.; Gandil, M.; Iwasaki, T.; Taniguchi, T.; Kieschnick, M.; Meijer, J.; Hatano, M.; Gali, A.; Becher, C. Spectroscopic Investigations of Negatively Charged Tin-Vacancy Centres in Diamond. *New J. Phys.* **2020**, *22*, 013048.

(23) Trusheim, M. E.; Pingault, B.; Wan, N. H.; Gündoğan, M.; De Santis, L.; Debroux, R.; Gangloff, D.; Purser, C.; Chen, K. C.; Walsh, M.; Rose, J. J.; Becker, J. N.; Lienhard, B.; Bersin, E.; Paradeisanos, I.; Wang, G.; Lyzwa, D.; Montblanch, A. R. P.; Malladi, G.; Bakhru, H.; Ferrari, A. C.; Walmsley, I. A.; Atatüre, M.; Englund, D. Transform-Limited Photons from a Coherent Tin-Vacancy Spin in Diamond. *Phys. Rev. Lett.* **2020**, *124*, 023602.

(24) Rugar, A. E.; Dory, C.; Aghaeimeibodi, S.; Lu, H.; Sun, S.; Mishra, S. D.; Shen, Z. X.; Melosh, N. A.; Vučković, J. Narrow-Linewidth Tin-Vacancy Centers in a Diamond Waveguide. *ACS Photonics* **2020**, *7*, 2356–2361.

(25) Rugar, A. E.; Aghaeimeibodi, S.; Riedel, D.; Dory, C.; Lu, H.; McQuade, P. J.; Shen, Z. X.; Melosh, N. A.; Vučković, J. Quantum Photonic Interface for Tin-Vacancy Centers in Diamond. *Phys. Rev. X* **2021**, *11*, 031021.

(26) Kuruma, K.; Pingault, B.; Chia, C.; Renaud, D.; Hoffmann, P.; Iwamoto, S.; Ronning, C.; Lončar, M. Coupling of a Single Tin-Vacancy Center to a Photonic Crystal Cavity in Diamond. *Appl. Phys. Lett.* **2021**, *118*, 230601.

(27) Martínez, J. A.; Parker, R. A.; Chen, K. C.; Purser, C. M.; Li, L.; Michaels, C. P.; Stramma, A. M.; Debroux, R.; Harris, I. B.; Appel, M. H.; Nichols, E. C.; Trusheim, M. E.; Gangloff, D. A.; Englund, D.; Atatüre, M. Photonic Indistinguishability of the Tin-Vacancy Center in Nanostructured Diamond. *arXiv:2206.15239* **2022**.

(28) Bernien, H.; Hensen, B.; Pfaff, W.; Koolstra, G.; Blok, M. S.; Robledo, L.; Taminiau, T. H.; Markham, M.; Twitchen, D. J.; Childress, L.; Hanson, R. Heralded Entanglement between Solid-State Qubits Separated by Three Metres. *Nature* **2013**, *497*, 86–90.

(29) Hong, C. K.; Ou, Z. Y.; Mandel, L. Measurement of Subpicosecond Time Intervals between Two Photons by Interference. *Phys. Rev. Lett.* **1987**, *59*, 2044–2046.

(30) Bouchard, F.; Sit, A.; Zhang, Y.; Fickler, R.; Miatto, F. M.; Yao, Y.; Sciarrino, F.; Karimi, E. Two-Photon Interference: The Hong-Ou-Mandel Effect. *Reports Prog. Phys.* **2021**, *84*, 012402.

(31) Diel, E. E.; Lichtman, J. W.; Richardson, D. S. Tutorial: Avoiding and Correcting Sample-Induced Spherical Aberration Artifacts in 3D Fluorescence Microscopy. *Nat. Protoc.* **2020**, *15*, 2773–2784.

(32) Görlitz, J.; Herrmann, D.; Fuchs, P.; Iwasaki, T.; Taniguchi, T.; Rogalla, D.; Hardeman, D.; Colard, P. O.; Markham, M.; Hatano, M.; Becher, C. Coherence of a Charge Stabilised Tin-Vacancy Spin in Diamond. *npj Quantum Inf.* **2022**, *8*, 45.

(33) Rugar, A. E.; Dory, C.; Sun, S.; Vučković, J. Characterization of Optical and Spin Properties of Single Tin-Vacancy Centers in Diamond Nanopillars. *Phys. Rev. B* **2019**, *99*, 205417.

(34) Dietrich, A.; Jahnke, K. D.; Binder, J. M.; Teraji, T.; Isoya, J.; Rogers, L. J.; Jelezko, F. Isotopically Varying Spectral Features of Silicon-Vacancy in Diamond. *New J. Phys.* **2014**, *16*, 113019.

(35) Ekimov, E. A.; Krivobok, V. S.; Lyapin, S. G.; Sherin, P. S.; Gavva, V. A.; Kondrin, M. V. Anharmonicity Effects in Impurity-Vacancy Centers in Diamond Revealed by Isotopic Shifts and Optical Measurements. *Phys. Rev. B* **2017**, *95*, 094113.

(36) Ekimov, E. A.; Lyapin, S. G.; Boldyrev, K. N.; Kondrin, M. V.; Khmelnitskiy, R.; Gavva, V. A.; Kotereva, T. V.; Popova, M. N. Germanium–Vacancy Color Center in Isotopically Enriched Diamonds Synthesized at High Pressures. *JETP Lett.* **2015**, *102*, 701–706.

(37) Londero, E.; Thiering, G.; Razinkovas, L.; Gali, A.; Alkauskas, A. Vibrational Modes of Negatively Charged Silicon-Vacancy Centers in Diamond from Ab Initio Calculations. *Phys. Rev. B* **2018**, *98*, 035306.

(38) Gali, A.; Maze, J. R. Ab Initio Study of the Split Silicon-Vacancy Defect in Diamond: Electronic Structure and Related Properties. *Phys. Rev. B* **2013**, *88*, 235205.

(39) NIST. *Atomic Weights and Isotopic Compositions for All Elements*. https://physics.nist.gov/cgi-bin/Compositions/stand_alone.pl?ele=&all=all&ascii=html&isotype=some.

(40) Rogers, L. J.; Jahnke, K. D.; Metsch, M. H.; Sipahigil, A.; Binder, J. M.; Teraji, T.; Sumiya, H.; Isoya, J.; Lukin, M. D.; Hemmer, P.; Jelezko, F. All-Optical Initialization, Readout, and Coherent Preparation of Single Silicon-Vacancy Spins in Diamond. *Phys. Rev. Lett.* **2014**, *113*, 263602.

(41) Abobeih, M. H.; Cramer, J.; Bakker, M. A.; Kalb, N.; Markham, M.; Twitchen, D. J.; Taminiau, T. H. One-Second Coherence for a Single Electron Spin Coupled to a Multi-Qubit Nuclear-Spin Environment. *Nat. Commun.* **2018**, *9*, 2552.

(42) Stas, P.; Huan, Y. Q.; Machielse, B.; Knall, E. N.; Knaut, C. M.; Assumpcao, D. R.; Sutula, M.; Wei, Y.; Ding, S. W.; Bhaskar, M. K.; Suleymanzade, A.; Pingault, B.; Sukachev, D. D.; Park, H.; Lonˇ, M.; Levonian, D. S.; Lukin, M. D. Robust Multi-Qubit Quantum Network Node with Integrated Error Detection. *arXiv:2207.13128* **2022**.

(43) Nagy, R.; Dasari, D. B. R.; Babin, C.; Liu, D.; Vorobyov, V.; Niethammer, M.; Widmann, M.; Linkewitz, T.; Gediz, I.; Stöhr, R.; Weber, H. B.; Ohshima, T.; Ghezellou, M.; Son,

N. T.; Ul-Hassan, J.; Kaiser, F.; Wrachtrup, J. Narrow Inhomogeneous Distribution of Spin-Active Emitters in Silicon Carbide. *Appl. Phys. Lett.* **2021**, *118*, 144003.

(44) Evans, R. E.; Sipahigil, A.; Sukachev, D. D.; Zibrov, A. S.; Lukin, M. D. Narrow-Linewidth Homogeneous Optical Emitters in Diamond Nanostructures via Silicon Ion Implantation. *Phys. Rev. Appl.* **2016**, *5*, 044010.

(45) Lang, J.; Häußler, S.; Fuhrmann, J.; Waltrich, R.; Laddha, S.; Scharpf, J.; Kubanek, A.; Naydenov, B.; Jelezko, F. Long Optical Coherence Times of Shallow-Implanted, Negatively Charged Silicon Vacancy Centers in Diamond. *Appl. Phys. Lett.* **2020**, *116*, 064001.

(46) Rogers, L. J.; Jahnke, K. D.; Teraji, T.; Marseglia, L.; Müller, C.; Naydenov, B.; Schauffert, H.; Kranz, C.; Isoya, J.; McGuinness, L. P.; Jelezko, F. Multiple Intrinsically Identical Single-Photon Emitters in the Solid State. *Nat. Commun.* **2014**, *5*, 4739.

(47) Haußler, S.; Hartung, L.; Fehler, K. G.; Antoniuk, L.; Kulikova, L. F.; Davydov, V. A.; Agafonov, V. N.; Jelezko, F.; Kubanek, A. Preparing Single SiV- Center in Nanodiamonds for External, Optical Coupling with Access to All Degrees of Freedom. *New J. Phys.* **2019**, *21*, 103047.

(48) Chen, D.; Froech, J.; Ru, S.; Cai, H.; Wang, N.; Adamo, G.; Scott, J.; Li, F.; Zheludev, N.; Aharonovich, I.; Gao, W. Quantum Interference of Resonance Fluorescence from Germanium-Vacancy Color Centers in Diamond. *Nano Lett.* **2022**, Article ASAP. DOI: 10.1021/acs.nanolett.2c01959.

(49) Sipahigil, A.; Goldman, M. L.; Togan, E.; Chu, Y.; Markham, M.; Twitchen, D. J.; Zibrov, A. S.; Kubanek, A.; Lukin, M. D. Quantum Interference of Single Photons from Remote Nitrogen-Vacancy Centers in Diamond. *Phys. Rev. Lett.* **2012**, *108*, 143601.

(50) Bernien, H.; Childress, L.; Robledo, L.; Markham, M.; Twitchen, D.; Hanson, R. Two-Photon Quantum Interference from Separate Nitrogen Vacancy Centers in Diamond. *Phys. Rev. Lett.* **2012**, *108*, 043604.

(51) Babinec, T. M.; Hausmann, B. J. M.; Khan, M.; Zhang, Y.; Maze, J. R.; Hemmer, P. R.; Lončar, M. A Diamond Nanowire Single-Photon Source. *Nat. Nanotechnol.* **2010**, *5*, 195–199.

(52) Nguyen, C. T.; Sukachev, D. D.; Bhaskar, M. K.; MacHielse, B.; Levonian, D. S.; Knall, E. N.; Stroganov, P.; Chia, C.; Burek, M. J.; Riedinger, R.; Park, H.; Lončar, M.; Lukin, M. D. An Integrated Nanophotonic Quantum Register Based on Silicon-Vacancy Spins in Diamond. *Phys. Rev. B* **2019**, *100*, 165428.

(53) Sohn, Y. I.; Meesala, S.; Pingault, B.; Atikian, H. A.; Holzgrafe, J.; Gündoğan, M.; Stavrakas, C.; Stanley, M. J.; Sipahigil, A.; Choi, J.; Zhang, M.; Pacheco, J. L.; Abraham, J.; Bielejec, E.; Lukin, M. D.; Atatüre, M.; Lončar, M. Controlling the Coherence of a Diamond Spin Qubit through Its Strain Environment. *Nat. Commun.* **2018**, *9*, 2012.

(54) Meesala, S.; Sohn, Y. I.; Pingault, B.; Shao, L.; Atikian, H. A.; Holzgrafe, J.; Gündoğan, M.; Stavrakas, C.; Sipahigil, A.; Chia, C.; Evans, R.; Burek, M. J.; Zhang, M.; Wu, L.; Pacheco, J. L.; Abraham, J.; Bielejec, E.; Lukin, M. D.; Atatüre, M.; Lončar, M. Strain Engineering of the Silicon-Vacancy Center in Diamond. *Phys. Rev. B* **2018**, *97*, 205444.

(55) Maity, S.; Shao, L.; Sohn, Y. I.; Meesala, S.; Machielse, B.; Bielejec, E.; Markham, M.; Lončar, M. Spectral Alignment of Single-Photon Emitters in Diamond Using Strain Gradient. *Phys. Rev. Appl.* **2018**, *10*, 024050.

(56) Wan, N. H.; Lu, T. J.; Chen, K. C.; Walsh, M. P.; Trusheim, M. E.; De Santis, L.; Bersin, E. A.; Harris, I. B.; Mouradian, S. L.; Christen, I. R.; Bielejec, E. S.; Englund, D. Large-Scale Integration of Artificial Atoms in Hybrid Photonic Circuits. *Nature* **2020**, *583*, 226–231.

(57) Aghaeimeibodi, S.; Riedel, D.; Rugar, A. E.; Dory, C.; Vučković, J. Electrical Tuning of Tin-Vacancy Centers in Diamond. *Phys. Rev. Appl.* **2021**, *15*, 064010.

(58) De Santis, L.; Trusheim, M. E.; Chen, K. C.; Englund, D. R. Investigation of the Stark Effect on a Centrosymmetric Quantum Emitter in Diamond. *Phys. Rev. Lett.* **2021**, *127*, 147402.

(59) Ziegler, J. F.; Ziegler, M. D.; Biersack, J. P. SRIM - The Stopping and Range of Ions in Matter (2010). *Nucl. Instruments Methods Phys. Res. Sect. B* **2010**, *268*, 1818–1823.

(60) Akaishi, M.; Kanda, H.; Yamaoka, S. High Pressure Synthesis of Diamond in the Systems of Grahpite-Sulfate and Graphite-Hydroxide. *Jpn. J. Appl. Phys.* **1990**, *29*, L1172–L1174.

(61) Binder, J. M.; Stark, A.; Tomek, N.; Scheuer, J.; Frank, F.; Jahnke, K. D.; Müller, C.; Schmitt, S.; Metsch, M. H.; Unden, T.; Gehring, T.; Huck, A.; Andersen, U. L.; Rogers, L. J.; Jelezko, F. Qudi: A Modular Python Suite for Experiment Control and Data Processing. *SoftwareX* **2017**, *6*, 85–90.

(62) Perdew, J. P.; Burke, K.; Ernzerhof, M. Generalized Gradient Approximation Made Simple. *Phys. Rev. Lett.* **1996**, *77*, 3865–3868.

(63) Troullier, N.; Martins, J. L. Efficient Pseudopotentials for Plane-Wave Calculations. *Phys. Rev.* **1991**, *43*, 1993–2006.

(64) Saito, M.; Oshiyama, A.; Sugino, O. Energetics and Local Vibrations of the DX Center in GaAs. *Phys. Rev. B* **1993**, *47*, 13205–13214.

# Supporting Information for

# Identical Photons from Multiple Tin-Vacancy

# Centers in Diamond

*Yasuyuki Narita[1], Peng Wang[1], Kazuki Oba[1], Yoshiyuki Miyamoto[2], Takashi Taniguchi[3], Shinobu Onoda[4], Mutsuko Hatano[1], Takayuki Iwasaki[1,\*]*

[1]Department of Electrical and Electronic Engineering, School of Engineering, Tokyo Institute of Technology, Meguro, Tokyo 152-8552, Japan

[2]Research Center for Computational Design of Advanced Functional Materials, National Institute of Advanced Industrial Science and Technology, Tsukuba, Ibaraki 305-8568, Japan

[3]International Center for Materials Nanoarchitectonics, National Institute for Materials Science, Tsukuba, Ibaraki 305-0044, Japan

[4]Takasaki Advanced Radiation Research Institute, National Institutes for Quantum Science and Technology, 1233 Watanuki, Takasaki, Gunma 370-1292, Japan

1. **Details of the computational scheme**

The first-principles calculation based on density functional theory (DFT) was performed for the tin-vacancy (SnV) split center in the -1 charged state by using the conjugated-gradient method.[1] The generalized gradient approximation (GGA) for the exchange correlation potentials, a $3 \times 3 \times 3$ supercell of the cubic diamond (216 atoms per cell without vacancy), and Γ point sampling in the momentum space were employed. For the GGA, the PBE functional[2] was chosen. Interactions between valence electrons and ions were treated by Troullier–Martins-type pseudopotentials.[3] A cutoff energy of 62 Ry was used for the plane-wave basis set to express the valance electron wavefunctions. A spin-unpolarized approximation was used since the contribution of the spin states is marginal for the force field in the diamond lattice and impurities.

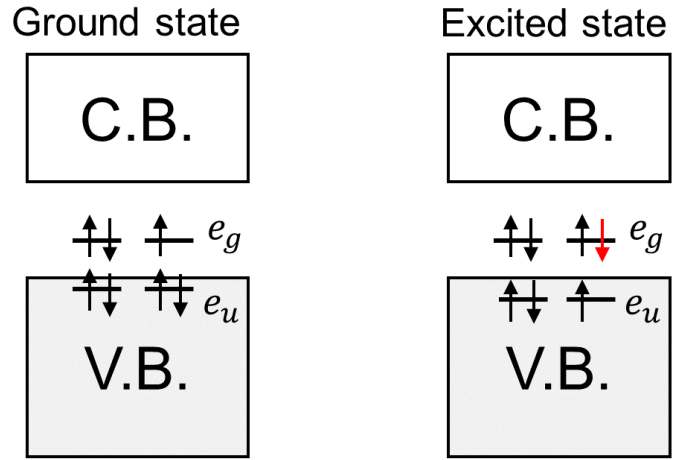

**Figure S1.** Energy diagram of the ground and excited states of the SnV center with the -1 charged state used for the force constant calculations. The valence and conduction bands are expressed by shaded and open square boxes, respectively. The red arrow on the right panel shows an optically excited electron. The bars denote energy levels, $e_g$ and $e_u$ states. Note that the $e_u$ level in the right panel is lowered.

The atomic configuration of the SnV center was optimized under the electronic ground and excited states. Both states yielded optimized geometries under the $D_{3d}$ symmetry. As illustrated in the left panel of Fig. S1, the top two occupied bands ($e_g$ states) are doubly degenerated and contain three electrons with the -1 charged state. Optical transition matrix elements between the $e_g$ states and the doubly degenerated states just below the valence band top ($e_u$ states) were recognized as nonzero. This is analogous to the fact reported for the SiV center.[4] Then, the optically excited state was approximated by promoting one electron occupation from the $e_u$ state to the $e_g$ state, as illustrated in the right panel of Fig. S1. Within the spin-unpolarized approximation, the ground state electron occupation was expressed by one electron per spin on the $e_u$ states and by 0.75 electrons per spin on the $e_g$ state. On the other hand, the excited state electron occupation was expressed by 0.75 electrons per spin on the $e_u$ state and one electron per spin on the $e_g$ state. Contrary to the SiV case,[4] this electron promotion accompanied by geometric optimization further lowered the $e_u$ level, as illustrated in the right panel of Fig. S1.

The zero-point lattice vibration energy was estimated within the harmonic approximation by taking the $A_{2u}$ mode (Sn replacement in the <111> direction) and doubly degenerated $E_u$ mode (Sn replacement in the <1$\bar{1}$0> or <$\bar{1}$10> directions) into account. These modes are the same as those of the SiV center.[5] The ZPL energy of the optical absorption of the SnV center was estimated within the harmonic approximation[5,6] as

$$E_{opt}^{gap} + \frac{\hbar}{2}\left[\Omega_{n_{S_n,A_{2u}}}^e + 2\Omega_{n_{S_n,E_u}}^e - \Omega_{n_{S_n,A_{2u}}}^g - 2\Omega_{n_{S_n,E_u}}^g\right]$$

Here, $E_{opt}^{gap}$ is the optical gap without the phonon contribution, and the second term is the phonon contribution. With the relation $\Omega_{n_{S_n}} = \sqrt{k/m_n}$, the energy becomes

$$E_{opt}^{gap} + \frac{\hbar}{2} \frac{1}{\sqrt{m_n}} \left[ \sqrt{k_{A_{2u}}^e} + 2\sqrt{k_{E_u}^e} - \sqrt{k_{A_{2u}}^g} - 2\sqrt{k_{E_u}^g} \right]$$

where $k$ and $m_n$ represent the force constant of the atomic displacement and mass of a $^n$Sn atom, respectively. The change in ZPL when $^n$Sn with a mass of $m_n$ is replaced by $^{n^*}$Sn with a mass of $m_{n^*}$ is given as

$$\Delta E_{n,n^*} = \frac{\hbar}{2} \left( \frac{1}{\sqrt{m_n}} - \frac{1}{\sqrt{m_{n^*}}} \right) \left( \sqrt{k_{A_{2u}}^e} + 2\sqrt{k_{E_u}^e} - \sqrt{k_{A_{2u}}^g} - 2\sqrt{k_{E_u}^g} \right)$$

The force field on the Sn atom with the displacement from its stable position by a 0.05 Bohr radius was calculated under the electronic excited/ground state in the <111> direction to obtain the force constants $k_{A_{2u}}^{e/g}$ and in the <1$\bar{1}$0> direction to obtain the force constants $k_{E_u}^{e/g}$. Table 1 shows the currently computed force constants in these modes under the electronic ground and excited states and the deduced vibration frequencies.

**Table 1.** Calculated force constants and vibration frequencies for Sn in the SnV center.

|  | Ground state | Excited state |
| --- | --- | --- |
| $k_{A_{2u}}$ (HR/Bohr$^2$) | 0.317201 | 0.386091 |
| $k_{E_u}$ (HR/Bohr$^2$) | 0.418695 | 0.435863 |
| $\Omega_{A_{2u}}$ (meV) | 32.1 | 35.9 |
| $\Omega_{E_u}$ (meV) | 37.7 | 38.4 |

2. **Depth of the SnV center**

Experimentally, the depth of the SnV was estimated from confocal fluorescence microscope (CFM) imaging (Fig. 1e in the main text). The mismatch of the refractive index between diamond and vacuum changes the depth of the CFM image compared to the real depth. Thus, we estimated the real depth of the SnV emitter from the CFM image by correcting according to the following relationship:[7]

$$\frac{D}{d} = \frac{\tan\left(\sin^{-1}\frac{0.5NA}{n_1}\right)}{\tan\left(\sin^{-1}\frac{0.5NA}{n_2}\right)}$$

where D is the actual laser focus point and d is the distance that the objective lens moves. $n_1$ and $n_2$ denote the refractive indices of vacuum (1.0) and diamond (2.4), respectively. Using an NA of 0.95, we obtained a correction factor (*D/d*) of 2.67. Therefore, the observed depth of ~1.2 μm corresponds to an actual emitter depth of ~3.2 μm, which agrees with the estimation by SRIM.[8]

3. **Other sets of identical photons from multiple SnV centers**

Figure S2 shows other measurements of identical photons from separated SnV centers at different resonant energies in Sample 1. In Fig. S2a, the two SnV centers have linewidths of 30 MHz (blue curve) and 35 MHz (red curve) with a center separation of ~4 MHz. Another measurement in Fig. S2b depicts the spectral overlap of three separated SnV centers, again with narrow linewidths of 34-37 MHz, showing that most of the peak areas overlap among the three emitters.

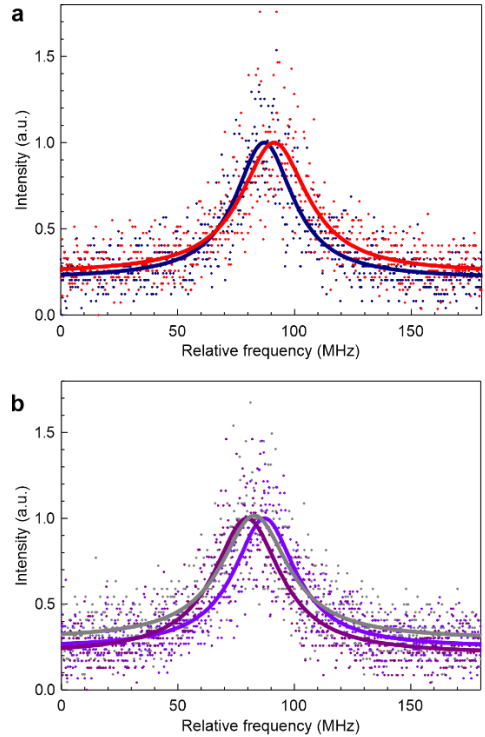

**Figure S2.** Identical photons from multiple separated SnV centers. Spectral overlap of (a) two and (b) three SnV centers. The experimental data are obtained with a single laser scan. The solid lines are fitted. The data and fitting are normalized with the maximum of the fitting for each curve. The frequency is given relative to 484.13085 THz in Panel (a) and 484.1312 THz in Panel (b).

4.  **Inhomogeneous distribution of the SnV centers in Sample 2**

Sample 2 was prepared with the same ion implantation and HPHT anneal procedures, but the surface of this sample was polished after the anneal. The HPHT anneal process modifies the surface morphology of diamond by etching and/or diamond growth. This hampers the fabrication of nanophotonic structures using high-quality SnV centers. To find a solution, here, we have tried the surface polishing of Sample 2. The polishing was done at a miscut angle of 0.5°, and SnV centers remained near the edge of the substrate after polishing were measured. As seen in Fig. S3, the surface of Sample 2 became flat compared with non-polished surface of Sample 1. All PLE measurements of Sample 2 were performed after the polishing. Figure S3 shows the inhomogeneous distribution of the resonant frequencies of 53 SnV centers. The resonant frequencies match the three regions observed in Sample 1, with the highest count in P1 for Sample 2 as well. A Gaussian fitting to the P1 region gives rise to a narrow FWHM of 5.2 GHz. Although this is slightly broader than that of Sample 1 (3.9 GHz), high-quality SnV centers are reproducibly fabricated.

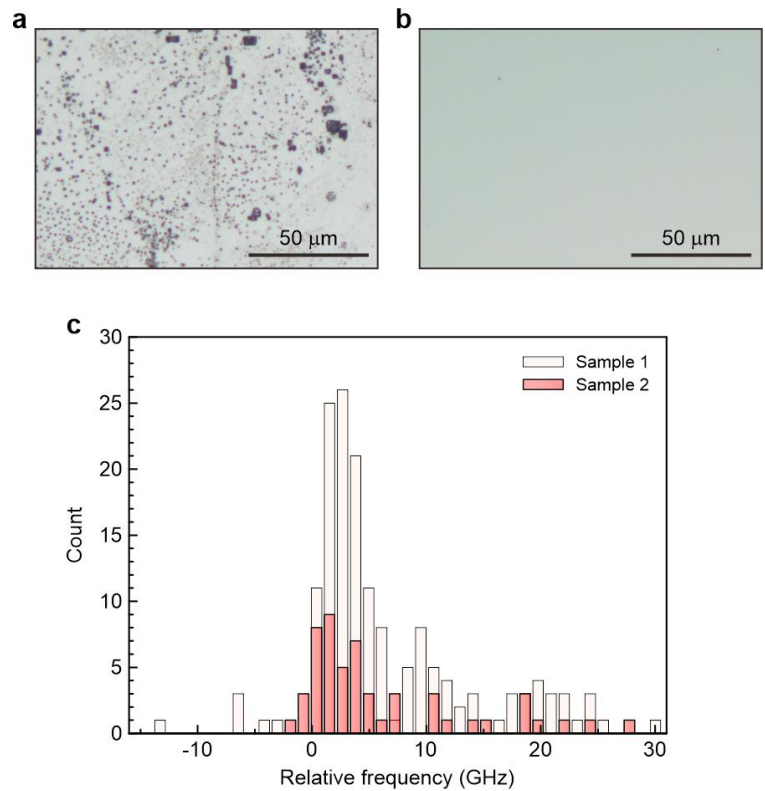

**Figure S3.** Optical microscope images of (a) Sample 1 and (b) Sample 2. The surface of Sample 2 was polished after high-temperature annealing. (c) Histogram of the resonant frequencies of SnV centers in Sample 2. The histogram of Sample 1 is also shown, same data as the main text. The two sets of data are shown overlaid. The frequency is the shift relative to 484.130 THz.

REFERENCES

(1) Saito, M.; Oshiyama, A.; Sugino, O. Energetics and Local Vibrations of the DX Center in GaAs. *Phys. Rev. B* **1993**, *47*, 13205–13214.

(2) Perdew, J. P.; Burke, K.; Ernzerhof, M. Generalized Gradient Approximation Made Simple. *Phys. Rev. Lett.* **1996**, *77*, 3865–3868.

(3) Troullier, N.; Martins, J. L. Efficient Pseudopotentials for Plane-Wave Calculations. *Phys. Rev. B* **1991**, *43*, 1993–2006.

(4) Gali, A.; Maze, J. R. Ab Initio Study of the Split Silicon-Vacancy Defect in Diamond: Electronic Structure and Related Properties. *Phys. Rev. B* **2013**, *88*, 235205.

(5) Londero, E.; Thiering, G.; Razinkovas, L.; Gali, A.; Alkauskas, A. Vibrational Modes of Negatively Charged Silicon-Vacancy Centers in Diamond from Ab Initio Calculations. *Phys. Rev. B* **2018**, *98*, 035306.

(6) Dietrich, A.; Jahnke, K. D.; Binder, J. M.; Teraji, T.; Isoya, J.; Rogers, L. J.; Jelezko, F. Isotopically Varying Spectral Features of Silicon-Vacancy in Diamond. *New J. Phys.* **2014**, *16*, 113019.

(7) Diel, E. E.; Lichtman, J. W.; Richardson, D. S. Tutorial: Avoiding and Correcting Sample-Induced Spherical Aberration Artifacts in 3D Fluorescence Microscopy. *Nat. Protoc.* **2020**, *15*, 2773–2784.

(8) Ziegler, J. F.; Ziegler, M. D.; Biersack, J. P. SRIM - The Stopping and Range of Ions in Matter (2010). *Nucl. Instruments Methods Phys. Res. Sect. B* **2010**, *268*, 1818–1823.